# ARTICLE

# Synthesis and Crystal Growth of Tetragonal $\beta$-Fe$_{1.00}$Se

**Cevriye Koz**[1], **Marcus Schmidt**[1], **Horst Borrmann**[1], **Ulrich Burkhardt**[1], **Sahana Rößler**[1], **Wilder Carrillo-Cabrera**[1], **Walter Schnelle**[1], **Ulrich Schwarz**\*[1], and **Yuri Grin**[1]

[1] Max-Planck-Institut für Chemische Physik fester Stoffe, Nöthnitzer Str. 40, 01187 Dresden, Germany

**Keywords:** Superconductivity; chemical vapor transport; selenium; iron

**Abstract.** Single crystals of tetragonal $\beta$-Fe$_{1.00}$Se were grown from polycrystalline material by chemical vapor transport reaction at temperatures below 723 K using AlCl$_3$ as transport additive. The plate-shaped single crystals have edge lengths up to 4 mm. The single crystals display complete diamagnetic shielding factor $4\pi\chi \approx -1$ below the superconducting transition.

## Introduction

Since the discovery of superconductivity in the iron-based compound LaO$_{1-x}$F$_x$FeAs with a critical temperature $T_c$ of 26 K,[1] a vast number of studies has been published on materials with similar layered crystal structures. The iron-based superconductors reported so far can be categorized into four major classes according to their structural organization. Among these different families, tetragonal $\beta$-Fe$_x$Se ($P4/nmm$) with a $T_c$ of 8 K can be considered as a reference material because of its archetypical structure pattern.[2] The phase diagram of the system Fe–Se indicates a homogeneity range of $\beta$-Fe$_x$Se from 51.0 to 50.6 $at.\%$ Se.[3] Later studies evidence that single phase samples of $\beta$-Fe$_x$Se are obtained only for compositions very close to 1:1[4] so that superconducting materials with nominal compositions FeSe$_{0.82}$[2] or FeSe$_{0.92}$[5] are clearly outside the homogeneity range of $\beta$-Fe$_x$Se.

The synthesis of $\beta$-Fe$_x$Se in polycrystalline or single crystalline form is a challenging task because of the peritectoid formation of the phase. Above 730 K, $\beta$-Fe$_x$Se decomposes into Fe-deficient $\delta$-Fe$_{1-y}$Se and $\alpha$-Fe [solid solution Fe(Se)].[6-8] Therefore, successful single crystal preparation requires a flux medium like LiCl/CsCl.[9] However, for single crystals manufactured with other alkali-halide fluxes [9–12] at temperatures around 973 K, impurities like $\delta$-Fe$_{1-y}$Se and $\alpha$-Fe are reported. Methods that rely on growth directly from the melt, like the Bridgman technique,[13] or vapor self-transport,[14] did not succeed for $\beta$-Fe$_x$Se, and even chemical vapor transport (CVT) studies[15-16] resulted in different phases within the same batch[17] or showed coexistence of $\delta$-Fe$_{1-y}$Se and $\beta$-Fe$_x$Se within individual crystals.[18]

The first part of the paper presents a detailed study of compositions close to the reported homogeneity range of $\beta$-Fe$_x$Se[4]. Here, nominal compositions between Fe$_{0.98}$Se and Fe$_{1.02}$Se are investigated. In the second part, we present results of the CVT technique to grow plate-shaped $\beta$-Fe$_{1.00}$Se single crystals at temperatures below 723 K. Finally, structural as well as physical properties of the obtained single crystals are reported.

## Experimental Section

**Sample preparation:** Polycrystalline samples were prepared by solid state reaction of iron pieces (Alfa Aesar, 99,995 %) with selenium shots (Chempur, 99,999 %) in substance amount fractions close to 1:1 (typically Fe:Se between 0.98 and 1.02). Glassy carbon crucibles with lids were filled with Fe/Se mixtures and placed in quartz ampoules which were sealed under vacuum. For the synthesis route, the procedure of an earlier study was optimized.[4] In order to ensure homogeneity, the raw product was ground, cold-pressed and annealed at 653 K for 2–5 days before being quenched in water to room temperature. Sample handling and preparation were performed in argon-filled glove boxes. Single crystals of $\beta$-Fe$_x$Se were grown from polycrystalline material by chemical vapor transport using anhydrous AlCl$_3$ (Alfa Aesar, 99.99 %) as transport additive. The addition of AlCl$_3$ is intrinsically no suitable means to raise the formation iron chloride transport species. However, reaction with traces of water, e.g., at the surface of the quartz ampoules, yields:

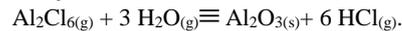
Al$_2$Cl$_{6(g)}$ + 3 H$_2$O$_{(g)}$ ⇌ Al$_2$O$_{3(s)}$ + 6 HCl$_{(g)}$.

The hydrogen chloride gas undergoes the consecutive reaction

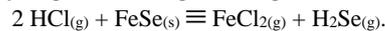
2 HCl$_{(g)}$ + FeSe$_{(s)}$ ⇌ FeCl$_{2(g)}$ + H$_2$Se$_{(g)}$.

In the presence of Al$_2$Cl$_6$, the formation of gas complexes might be relevant for chemical transport:[15]

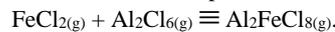
FeCl$_{2(g)}$ + Al$_2$Cl$_{6(g)}$ ⇌ Al$_2$FeCl$_{8(g)}$.

In case of the chemical transport yielding $\beta$-Fe$_x$Se, the commonly used HCl source NH$_4$Cl is not suited since the partial pressure of the hydrogen chloride is insufficient at temperatures below 700 K. For transport experiments, evacuated, sealed quartz ampoules (diameter 20 mm, length 100 mm) were filled with mixtures of typically 1 g of FeSe powder and ~ 20 mg of AlCl$_3$ before being placed horizontally inside a two-zone furnace at temperatures from $T_2$ = 673 K to $T_1$ = 573 K. Typically, crystals were grown for 2 months. Finally, the ampoule was quenched in water. The product, which contains plate-shaped single crystals with edge lengths up to 400 μm perpendicular to the $c$–axis, was washed with ethanol several times to remove remaining condensed gas phase, dried under vacuum and stored in argon-filled glove boxes. By extending the growth time to one year, larger single crystals with dimensions up to $4 \times 2 \times 0.03$ mm$^3$ were grown.

**X-ray and electron diffraction:** Samples were investigated by X-ray powder diffraction (Huber Image Plate Camera G670) using Co K$\alpha_1$ radiation ($\lambda$ = 1.788965 Å). The lattice parameters of samples were calculated with TiO$_2$ (rutile, $a$ = 4.59393(4) Å, $c$ = 2.95887(3) Å) as an internal standard using the program package WinCSD.[19] Single crystal X-ray diffraction data were collected with a Rigaku AFC7 diffractometer equipped with a Saturn 724+ CCD detector using Mo K$\alpha$ ($\lambda$ = 0.71073 Å) radiation. The SHELX-97 program package was used for structure refinements.[20] High- and low-temperature powder X-ray diffraction data were collected at the high-resolution powder diffraction beamline ID31 ($\lambda$ = 0.39491 Å and $\lambda$ = 0.39996 Å, respectively) of the ESRF. High-temperature measurements up to 873 K were realized with a hot-air blower adapted to the diffraction setup while a liquid-He flow cryostat was used at low-temperatures down to 40 K. Lattice parameter determinations and structure refinements were performed by least-squares methods using complete diffraction profiles and the program FullProf.[21] In the Rietveld refinement procedures, the March-Dollase approach to depict the preferred orientation was complemented by models for describing the anisotropic peak broadening.[22-23] For selected area electron diffraction (SAED) investigations, focused ion beam (FIB) thin cuts of single crystals were prepared by means of a FEI Quanta 200 3D dual beam device (FEI Company, Eindhoven, NL). Conventional transmission electron microscopy (TEM) and selected area diffraction were performed by FEI TECHNAI 10 (100 kV) microscope, equipped with a 2k CCD camera (TemCam-F224HD, TVIPS).

**Chemical composition and thermal behavior:** The synthesized ingots and single crystals were characterized by wavelength dispersive X-ray spectroscopy on an electron microprobe (WDXS, CAMECA SX 100, with FeGe and Se as standards). Chemical analysis by the inductively coupled plasma method (ICP-OES, Varian, VISTA RL) reveal that oxygen and carbon impurities are below the detection limit of 0.05 mass % and 0.06 mass %, respectively.

Differential thermal analysis (DSC) was performed with a Netzsch instrument DSC 404 C in the temperature range from 298 to 823 K with heating rates between 2 and 5 K/min. The polycrystalline materials for investigation were sealed in special quartz crucibles under vacuum.

**Physical property measurements:** Specific heat, $C_P(T)$, was measured using a Quantum Design physical properties measurement system (PPMS) with a heat-pulse relaxation technique. The electrical resistivity, $\rho(T)$, measurements were carried out with the same device using the standard four-probe method from 2 K to room temperature at zero field. The magnetic susceptibility, $\chi(T)$, was obtained by means of a SQUID magnetometer in magnetic fields of 20 Oe. The amount of $\alpha$-Fe impurities (actually Fe(Se) solid solution) was estimated from isothermal magnetization loops at 300 K in fields from −10 to 10 kOe assuming the saturation magnetization of $\alpha$-Fe of 217.6 emu/g at 10 kOe and 300 K.[24] The magnetization curves starting from 10 kOe were extrapolated to zero and the magnetization intercepts were divided by 217.6 emu/g. These ferromagnetic magnetizations yield estimates for the substance amount fraction of $\alpha$-Fe.

## Results and Discussion

X-ray powder diffraction experiments exhibit that the prepared polycrystalline samples contain mainly $\beta$-Fe$_x$Se. Before annealing, both hexagonal Fe$_7$Se$_8$ and $\delta$-Fe$_{1-y}$Se are present besides unreacted iron (Fig.1a). Within the nominal composition range from Fe$_{1.02}$Se to Fe$_{0.99}$Se, the reflections

**Figure 1.** X-ray powder diffraction diagrams of samples with nominal ratios Fe:Se between 0.98 and 1.02 (a) before and (b) after annealing at 653 K for 4 days. Lines of impurity phases are marked by 'H' when hexagonal Fe$_7$Se$_8$ and $\delta$-Fe$_{1-y}$Se coexist, those of Fe$_7$Se$_8$ are designated by 'h'. "H+T" or "h+T" refers to a peak of $\beta$-Fe$_x$Se overlapping with lines of the impurity phases. The diffraction patterns are recorded using Co K$\alpha_1$ radiation.

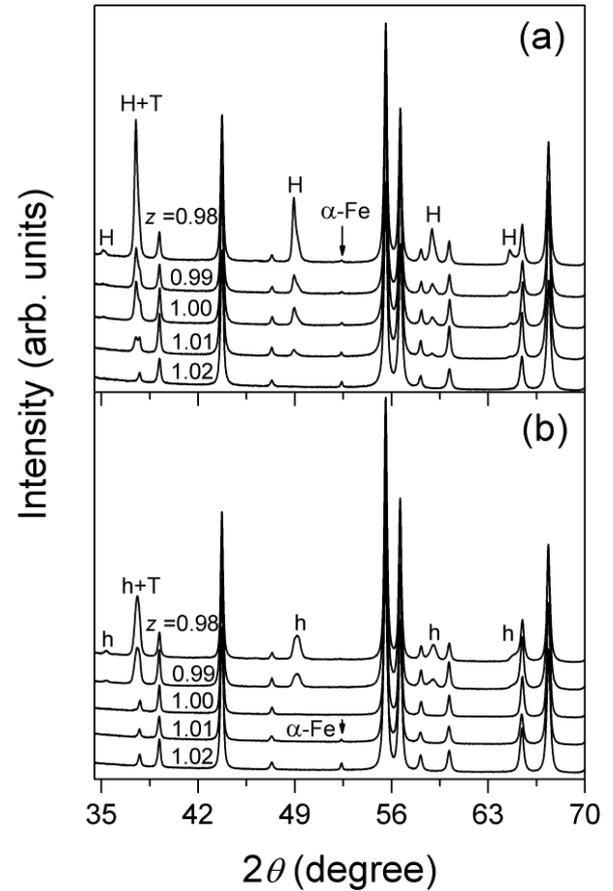

Fe$_{1-y}$Se disappear after annealing. However unreacted iron is still observed in samples with nominal Fe excess while a surplus of Se induces the formation of hexagonal Fe$_7$Se$_8$ (Fig. 1b). Figure 2 shows the unit cell parameters of $\beta$-Fe$_x$Se for Fe:Se ratios between 0.98 and 1.02. Two series of compositions are prepared independently at the same conditions in order to guarantee reproducible results. The analysis narrows the nominal composition range of $\beta$-Fe$_x$Se to 1.000(5) $\leq x \leq$ 1.010(5). According to WDXS and ICP analyses, polycrystalline single phase samples $\beta$-Fe$_x$Se with a nominal composition $x$ = 1 exhibit parameter values 0.98(1) $\leq x \leq$ 1.01(2).

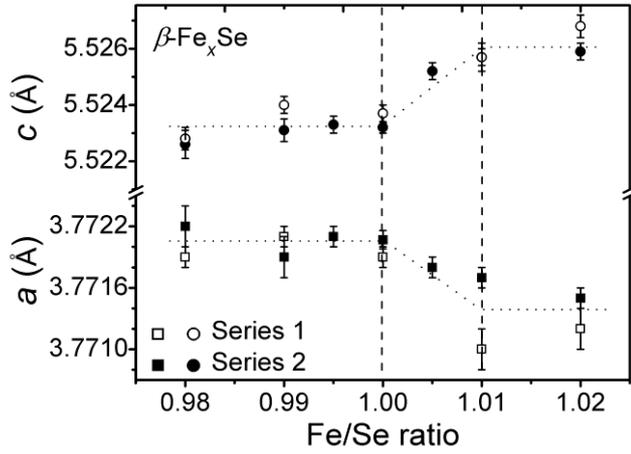

**Figure 2.** Lattice parameters of $\beta$-Fe$_x$Se at room temperature for the nominal compositions Fe:Se between 0.98 and 1.02.

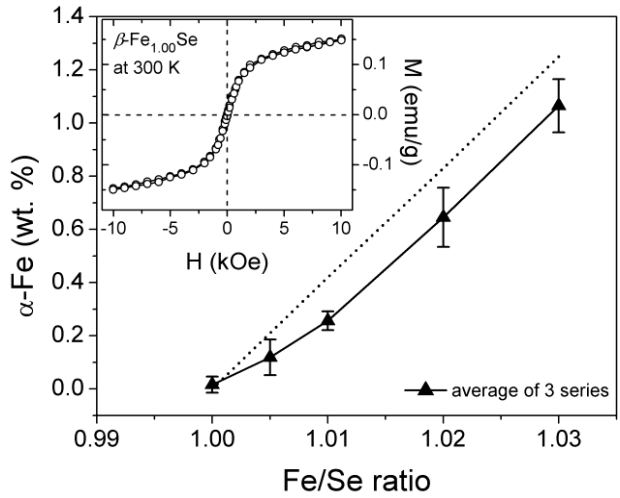

**Figure 3.** Content of $\alpha$-Fe in polycrystalline samples determined from magnetization measurements of three different series at room temperature for nominal ratios Fe:Se between 1.00 and 1.03. Error bars represent 3σ standard deviations. The dotted line represents the equivalent amount of $\alpha$-Fe assuming that the phase $\beta$-Fe$_x$Se adopts exactly $x = 1$. The inset shows the magnetization of a polycrystalline sample $\beta$-Fe$_{1.00}$Se as a function of the applied field at $T$ = 300 K.

In addition to these techniques, the amount of elemental iron in the samples is rated by magnetization measurements (Figure 3 inset, for details of the measurement procedure see Experimental). According to these estimates, the best polycrystalline samples contain between 100 and 300 ppm elemental iron. The determined equivalent amount of $\alpha$-Fe is shown in Figure 3 as a function of the nominal composition. We assign this ferromagnetic signal for the most part to unreacted $\alpha$-Fe. The magnetic signal in samples with Fe:Se ratios smaller than 1.00 is attributed to ferromagnetic contributions from Fe$_7$Se$_8$.[25]

Regarding the transformation of the 1:1 phase in the Fe−Se system, there are discussions concerning the involved phases. In order to gain insight into the relation between superconducting tetragonal $\beta$-Fe$_x$Se and $\delta$-Fe$_{1-y}$Se, findings of DSC and *in-situ* XRD studies are combined. In the DSC measurements (see Fig. S1 in the Supporting Information), an endothermic effect with onset at 733 K is attributed to the decomposition of $\beta$-Fe$_{1.00}$Se. Upon cooling, the reverse reaction is monitored at around 677 K. Although the majority of sample transforms back to $\beta$-Fe$_x$Se upon cooling, there remain small amounts of unreacted $\alpha$-Fe [iron-rich solid solution Fe(Se)] [8] and $\delta$-Fe$_{1-y}$Se. In agreement with the findings from the DSC experiments, X-ray diffraction data indicate that $\beta$-Fe$_{1.00}$Se remains stable up to 714 K upon heating (Figure 4). The decomposition of $\beta$-Fe$_{1.00}$Se into Fe-deficient $\delta$-Fe$_{1-y}$Se and traces of $\alpha$-Fe in full accordance with earlier reports.[6,26-27]

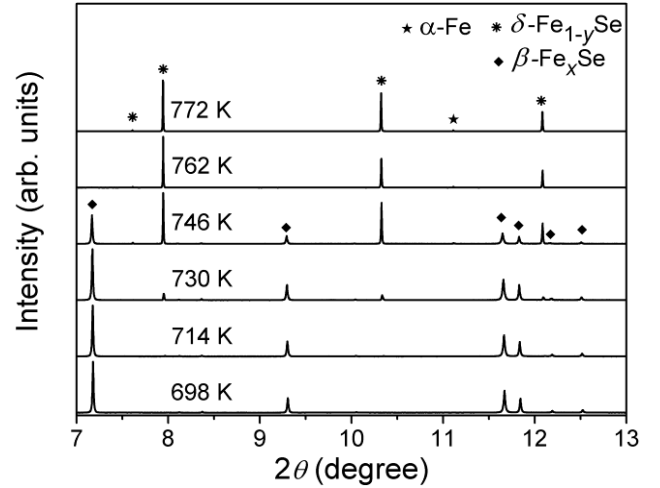

**Figure 4.** Powder X-ray diffraction diagrams at elevated temperatures in the range of the decomposition of $\beta$-Fe$_{1.00}$Se into $\delta$-Fe$_{1-y}$Se and $\alpha$-Fe [Fe(Se) solid solution]. The diffraction patterns are recorded in direction of increasing temperature using synchrotron radiation with λ = 0.39491 Å.

The refined XRD diagrams of samples at 673 K and 773 K (see Table 1 and Figures S2a and S2b in the Supporting Information) yield a composition of 50% Se in the tetragonal phase at 673 K. The finding of 51.1(3) at.% Se (Fe$_{0.96(1)}$Se) at 773 K for $\delta$-Fe$_{1-y}$Se is in agreement with earlier investigations.[27] The *in-situ* XRD measurements and the thermal analysis data give no indication for a reported[4] phase transition from tetragonal to hexagonal below 573 K.

**Table 1.** Crystallographic data based on refinements of powder X-ray diffraction data of $\beta$-Fe$_{1.00}$Se at 673 K and $\delta$-Fe$_{0.96}$Se at 773 K.

| Temperature | 673 K | 773 K |
|---|---|---|
| Compound | $\beta$-Fe$_{1.00}$Se | $\delta$-Fe$_{0.96(1)}$Se |
| Space group | $P4/nmm$ | $P6_3/mmc$ |
| a | 3.8280(2) Å | 3.7565(1) Å |
| c | 5.5821(3) Å | 5.9627(1) Å |
| γ | 90° | 120° |
| R$_{bragg}$; $\chi^2$ | 3.78; 2.50 | 2.75; 1.81 |
| R$_{wp}$; R$_{exp}$ | 9.28; 5.87 | 8.84; 6.56 |
| Atomic parameters | | |
| Fe | 2a (¾, ¼, 0) | 2a (0, 0, 0) |
| | B$_{iso}$ = 2.04(2) Å$^2$ | B$_{iso}$ = 3.80(4) Å$^2$ |
| | Occ. = 1 | Occ. = 0.96(1) |
| Se | 2c (¼, ¼, z) | 2c (⅓, ⅔, ¼) |
| | z = 0.2676(1) | |
| | B$_{iso}$ = 2.03(1) | B$_{iso}$ = 2.25(2) |
| | Occ. = 1 | Occ. = 1 |
| Weight fraction | 100 % | 98.8(5) % |

Figure 5 shows the temperature dependence of the magnetic susceptibility of polycrystalline samples with different nominal compositions upon zero-field-cooling (ZFC) and field-cooling (FC) in a magnetic field of 20 Oe. The characteristic changes indicate superconducting behaviour at low temperatures. Positive values in the non-superconducting range of samples with a slight Fe excess are consistent with magnetic impurities attributed to elemental iron. Besides, the $T_c$ onset slightly decreases with increasing equivalent α-Fe content and the highest $T_c$ of powder samples is observed at 8.2 K for a nominal composition $Fe_{1.00}$Se which is in good agreement with a previous studies on $\beta$-$Fe_x$Se.[2,4]

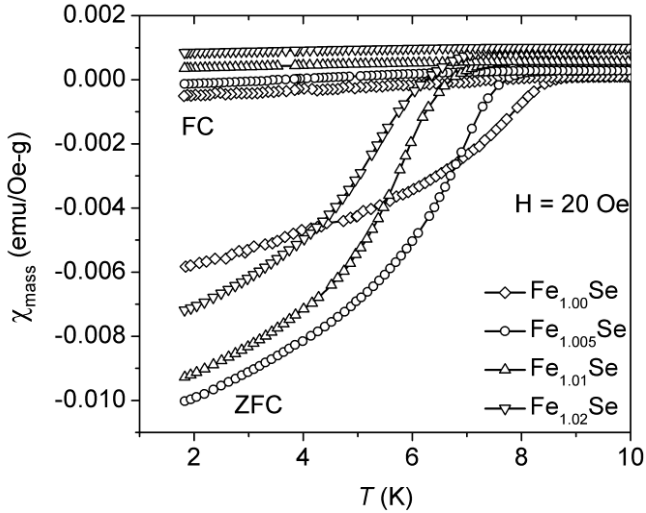

**Figure 5.** Magnetic susceptibility as a function of temperature for polycrystalline samples (nominal ratios Fe:Se from 1.00 to 1.02) in a magnetic field of 20 Oe. Both shielding and Meissner diamagnetic characteristics, as measured by zero field cooling (ZFC) and field cooling (FC) cycles, are shown.

Figure 6a shows a single crystal of $\beta$-$Fe_x$Se grown by a CVT reaction. The edge lengths perpendicular to the $c$–axis range typically from 400 μm after 2 months up to 4 mm after one year, the thickness in direction of the crystallographic $c$–axis usually amounts to some tens of micrometers.

Figure 6b shows the backscattering-electron image. Although some single crystals contain small segregations of $AlCl_3$ at the surface, the chlorine content in the bulk material is below the detection limit of 0.6 *wt.%*. According to the WDXS analysis, the composition of the single crystals corresponds to $Fe_{0.992(6)}$Se which agrees within uncertainty to the nominal composition of $Fe_{1.00}$Se.

Backscattering Laue diffraction measurements along the [001] direction evidence the single-crystalline nature even of the largest grown individuals (Figure 6c). The symmetry of the projection is *p4mm*. Selected area electron diffraction along the [100] zone axis exhibit Patterson symmetry *p2mm* (Figure 6d). The weaker but significant intensity of the (100) reflection is attributed to multiple scattering. Thus, the symmetry of the projections is consistent with the selected space group of the crystal structure. The observed plate-like morphology of $\beta$-$Fe_x$Se crystals and the obtained diffraction data are in accordance with the layered atomic pattern[28] implying that the compound grows preferentially perpendicular to the $c$–axis as observed in other layered chalcogenides.[29-30] According to the results of crystal structure refinements, large crystals exhibit pronounced extinction. Thus, the given results (Tables 2 and 3) are obtained for a crystal with dimensions $20 \times 15 \times 20$ μm$^3$. In the crystal structure of $\beta$-$Fe_{1.00}$Se each iron atom is tetrahedrally surrounded by four selenium atoms adopting Fe–Se distances of 2.3933(4) Å and Se–Fe–Se angles of 104.00(2) and 112.27(1) degrees, respectively.

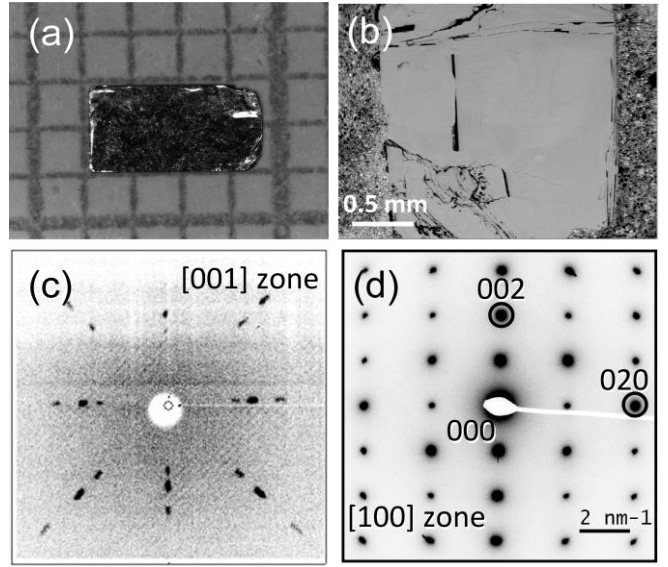

**Figure 6.** (a) $\beta$-$Fe_{1.00}$Se single crystal with dimensions $3.5 \times 1.9 \times 0.03$ mm$^3$, (b) backscattering electron (BSE) image of a single crystal, (c) backscattering Laue diffraction pattern along the $c$–axis (direction [001]), (d) selected area electron diffraction pattern along the $a$–axis (zone axis [100]).

**Table 2.** Crystallographic data of $\beta$-$Fe_{1.00}$Se refined from single crystal X-ray diffraction data.

| Space group | P4/nmm (no. 129) |
| --- | --- |
| $a; c$ / (Å) | 3.7719(1); 5.5237(3) [a] |
| Unit cell volume / (Å$^3$) | 78.587(5) |
| Z; $\rho_{calc}$ / (g cm$^{-3}$) | 2; 5.697 |
| T / (K) | 295 |
| $\theta$ range (°) | 3.69 to 30.94 |
| Indexes ranges | $-5 \leq h \leq 5$ |
| | $-5 \leq k \leq 5$ |
| | $-7 \leq l \leq 3$ |
| $\mu$ / (mm$^{-1}$) | 32.07 |
| F(000) / (e) | 120 |
| Absorption correction | Multi-scan |
| Reflections collected | 1155 |
| $R_{int}$, Independent reflections | 0.045, 95 |
| Refinement method | Full-matrix least-squares on $F^2$ |
| Refined parameters | 7 |
| Residuals [I > 2σ(I)] | R1 = 0.022, |
| | wR2 = 0.047 |
| Residuals (all data) | R1 = 0.022, |
| | wR2 = 0.047 |
| Goodness-of-fit on $F^2$ | 1.110 |
| Extinction coefficient | 0.053(11) |
| Largest diff. peak and hole / (e / Å$^3$) | 0.661 and -0.643 |

[a] Lattice parameters were calculated from X-ray powder diffraction data using $TiO_2$ as an internal standard.

**Table 3.** Atomic coordinates and displacement parameters[a] (given in $10^{-2}\text{Å}^2$) for $\beta$-Fe$_{1.00}$Se. Fe is located at $2a$ and Se at $2c$.

| Atom | $x$ | $y$ | $z$ | $U_{11}$ | $U_{33}$ | $U_{eq}$ |
|------|-----|-----|-----|----------|----------|----------|
| Fe | 3/4 | 1/4 | 0 | 0.96(4) | 1.66(5) | 1.19(3) |
| Se | 1/4 | 1/4 | 0.2667(1) | 1.33(3) | 1.26(4) | 1.31(3) |

[a] $U_{22} = U_{11}$, $U_{23} = U_{13} = U_{12} = 0$, $U_{eq} = 1/3\,(U_{11} + U_{22} + U_{33})$

These FeSe$_{4/4}$ tetrahedra condense via sharing of edges into infinite layers. Concerning the anisotropy of the atomic displacement, the ratio $U_{33}/U_{11}$ amounts to 1 for Se and 1.7 for Fe. Similar elongations of the displacement ellipsoids along the $c$–axis are reported for refinements of synchrotron radiation diffraction data on powder of $\beta$-Fe$_x$Se.[4,9] $\beta$-Fe$_x$Se single crystals are very soft and can easily be deformed during handling. The results of a detailed TEM investigation on deformed single crystals reveal that these crystals exhibit shear planes perpendicular to the $c$–axis. Consequently, stacking faults along the $c$–axis may substantially contribute to the large $U_{33}/U_{11}$ ratio of Fe. Variations of the site occupancy factor of iron or selenium did not yield significant differences of the residuals. This finding is in conformity with the results of the ICP and WDXS analyses which yield the composition Fe$_{1.00}$Se within experimental uncertainty.

Figure 7 presents the magnetization measurements with zero-field cooled and field-cooled protocols for $\beta$-Fe$_x$Se single crystals in a field of 20 Oe. The shape of the investigated single crystal was plate-like and the direction of the applied field is parallel to its long axis (field is in $a$–$b$ plane). Therefore, the demagnetization factor was assumed as zero. The superconducting transition is observed at $T_c^{onset}$ = 9 K in the magnetic susceptibility measurements, which is ≈ 1 K higher than the $T_c^{onset}$ of polycrystalline $\beta$-Fe$_x$Se samples. No magnetic discontinuity is observed between $T_c$ and 300 K. Thus, anomalies reported between 70 and 120 K[2, 32-34] are likely due to the presence of impurity phases such as Fe$_7$Se$_8$, Fe or iron oxide.

magnetization loops of single crystals at 300 K evidence a very low iron impurity level of ~ 4 ppm (not shown).

The temperature dependence of the electrical resistivity of a $\beta$-Fe$_x$Se single crystal is shown in Figure 8. The resistivity data exhibit metallic behaviour and a clear anomaly at arround 90 K, where a phase transition from tetragonal ($P4/nmm$) to orthorhombic ($Cmma$) is reported.[35] Results of X-ray powder diffraction experiments on polycrystalline $\beta$-Fe$_x$Se at low temperatures confirm that the anomaly is associated to the symmetry-breaking phase transition (Fig. 9 and Fig. S3 of the Supporting Information).

The specific heat of a $\beta$-Fe$_x$Se single crystal as a function of temperature is displayed in the upper inset of Figure 8.

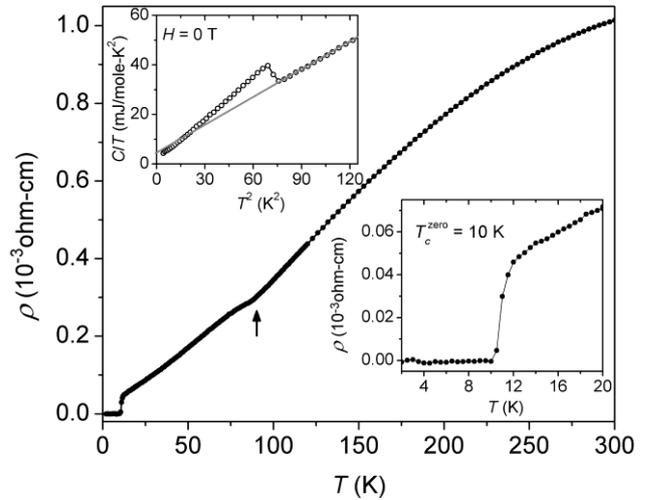

**Figure 8.** Resistivity of a $\beta$-Fe$_{1.00}$Se single crystal as a function of temperature. The kink at ≈ 90 K which is marked by an arrow is attributed to the phase transition from tetragonal to orthorhombic symmetry. The lower inset displays the resistivity below 20 K including the superconducting transition. The upper inset shows the specific heat as a function of temperature of a $\beta$-Fe$_{1.00}$Se single crystal. The grey solid line is the curve fitting the phonon and electronic contributions to the specific heat data.

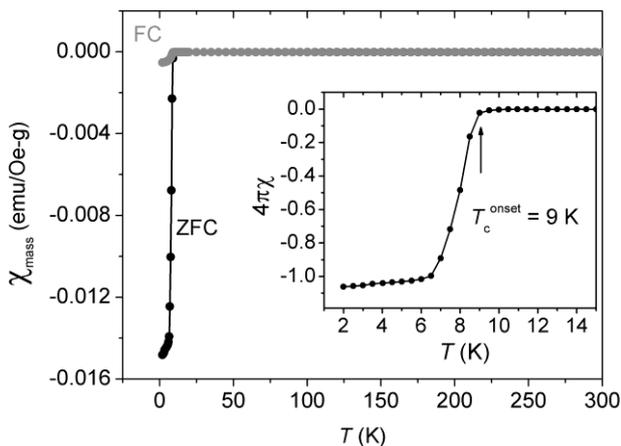

**Figure 7.** Magnetic susceptibility as a function of temperature for single crystalline $\beta$-Fe$_{1.00}$Se in an external field of $H_{ext}$ = 20 Oe. The inset shows volume susceptibility as a function of temperature.

The inset of Figure 7 shows the zero-field cooled dc-susceptibility curve below 15 K. The fraction of the volume that is screened by the superconducting currents, estimated from the dimensionless dc-susceptibility, yields almost the full screening value, $4\pi\chi = -1$. Besides, isothermal

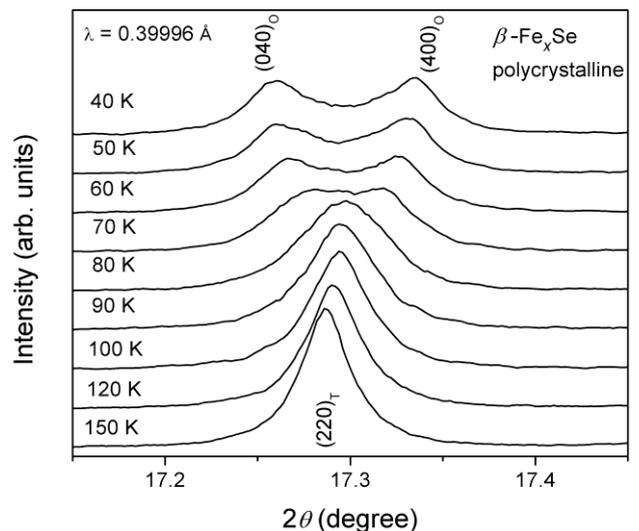

**Figure 9.** Powder X-ray diffraction diagrams of $\beta$-Fe$_{1.00}$Se at low temperatures. The peak splitting of $(220)_T$ at $2\theta \sim 17.29°$ into $(040)_O$ and $(400)_O$ below 90 K is consistent with the reported temperature-induced symmetry-breaking phase transition involving changes of the identity periods: $a_O = \sqrt{2}a_T$, $b_O = \sqrt{2}a_T$, $c_O = c_T$ (see also Tab. S1 in the Supporting Information).

The sizable jump associated to the superconducting transition indicates bulk nature of the superconductivity. The superconducting transition temperature 8.52(25) K and the jump at $T_c$ were determined by a graphical construction keeping the entropy balanced at the critical temperature. The insignificant broadening of the step shows that the sample is chemically homogeneous. Measurements at $H$ = 9 T (not shown) allow for an experimental determination of the normal state specific heat $C_n(T)$. Between 5 K and 13 K, $C_n(T)$ can be described by the simple equation $C_n(T) = \gamma_n T + \beta T^3$. In the normal state, the electronic coefficient of specific heat is $\gamma_n$ = 5.11(11) mJ mol$^{-1}$ K$^{-2}$ and the coefficient of the lattice contribution is $\beta$ = 0.372(1) mJ mol$^{-1}$ K$^{-4}$. The dimensionless specific-heat jump at $T_c$ is $\Delta C/\gamma_n T_c$ = 2.0(1). This value is significantly higher than the BCS value of 1.43 for the weak electron-phonon coupling scenario, but consistent with the recently proposed mechanism for the superconductivity in undoped FeSe.[36] The temperature dependence of the resistivity below 20 K is shown in the lower inset of Figure 8. A sharp drop in resistivity is observed below the $T_c^{onset}$ of 12 K and zero resistivity is monitored at 10 K. The determined residual resistivity ratio, RRR = $\rho$(300 K)/$\rho$(12 K) amounts to 22, which is the highest value reported for $\beta$-Fe$_x$Se so far. The residual resistivity ratio depends on the amount of impurities and lattice defects of the materials, so the high value of RRR evidences that single crystals grown by CVT have high purity and low defect concentration.

**Conclusions**

The phase $\beta$-Fe$_x$Se forms in the narrow homogeneity range 1.000(5) ≤ $x$ ≤ 1.010(5). The phase is stable up to roughly 733 K before a peritectoid decomposition into $\delta$-Fe$_{0.96(1)}$Se and solid solution Fe(Se) occurs. Chemical vapor transport reaction with AlCl$_3$ yields single crystals of tetragonal $\beta$-Fe$_{1.00}$Se at temperatures below 723 K. Although the CVT technique is a slow process which requires long growth times for $\beta$-Fe$_x$Se, it is possible to synthesize high-quality plate-shaped single crystals with up to 4 mm edge lengths. Magnetization measurements indicate that the crystals contain only 4 ppm magnetic impurities. The magnetic susceptibility and specific heat measurements evidence bulk superconductivity in $\beta$-Fe$_{1.00}$Se single crystals with onset temperatures up to 10 K.

**Supporting Information** (see footnote on the first page of this article) Differential Scanning Calorimetry Data of β-Fe$_{1.00}$Se as well as X-ray powder diffraction data of $\beta$-Fe$_{1.00}$Se and the products after peritectoid decomposition. Refinement results of the orthorhombic low-temperature modification of Fe$_{1.00}$Se are shown and the essential crystallographic data are listed. Cif-files of refined crystal structures have been deposited at the Cambridge Crystallographic Data Centre with the assigned numbers CCDC 977900-977903.

**Acknowledgement**

Stimulating discussions with L. H. Tjeng, S. Wirth and Y. Prots are gratefully acknowledged. The authors wish to thank G. Auffermann, U. Schmidt, A. Völzke, S. Kostmann, M. Eckert, S. Scharsach, R. Koban, A. A. Tsirlin, A. Henschel and S. Hückmann for their support and technical assistance. We also acknowledge the ESRF for granting beam time at ID31. The project is supported by the DFG SPP 1458.

**Entry for the Table of Contents**

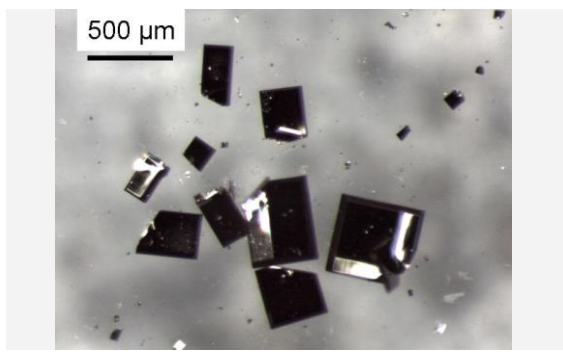

C. Koz, M. Schmidt, H. Borrmann, U. Burkhardt, S. Röβler, W. Carrillo-Cabrera, W. Schnelle, U. Schwarz*, and Yu. Grin

Synthesis and crystal growth of tetragonal $\beta$-Fe$_{1.00}$Se